\documentstyle[emulateapj]{article}
\input epsf.tex
\newcommand{\be}{\begin{equation} }
\newcommand{\ee}{\end{equation}}
\begin{document}
\title{The Origin of Primordial Dwarf Stars and Baryonic Dark Matter }
\author{Brad M. S. Hansen }
\affil{Canadian Institute for Theoretical Astrophysics, University of
Toronto, Toronto, ON M5S 3H8, Canada}
\authoremail{hansen@cita.utoronto.ca}

\begin{abstract}
I present a scenario for the production of low mass, degenerate dwarfs of mass $>0.1 M_{\odot}$ via the mechanism of
Lenzuni, Chernoff \& Salpeter (1992). Such objects meet the mass limit requirements for halo
dark matter from microlensing surveys while circumventing the chemical evolution constraints
on normal white dwarf stars. I describe methods to observationally constrain this scenario and suggest that
 such objects may originate in small clusters formed from the thermal instability of shocked, heated gas in dark matter haloes, such
as suggested by Fall \& Rees (1985) for globular clusters.
\end{abstract}

\keywords{accretion -- stars: brown dwarfs -- stars: formation --  Galaxy: halo -- cooling flows -- galaxies: formation
}

\section{Introduction}

The detection of microlensing  in the direction of the Large Magellanic Cloud  (Alcock et al 1993; Aubourg et al 1993)
has stimulated investigations into the nature of baryonic dark matter. In particular, the
lack of short timescale events (Alcock et al 1997; Aubourg et al 1995) appears to
 favour objects more massive than brown dwarfs. 
This mass limit, combined with direct observational constraints, favours white dwarfs more than main sequence stars (Bahcall et al 1994;
Graff \& Freese 1996; Hansen 1998).
However the problems associated
with the chemical pollution from prior evolutionary stages (e.g. Gibson \& Mould 1997) makes attempts to salvage 
the brown dwarf hypothesis worthwhile. Some have invoked dark clusters of brown dwarfs (Carr \& Lacey 1987; 
Kerins 1997, De Paolis et al 1998)
or spatially varying mass functions (Kerins \& Evans 1998) to explain the lack of short timescale events.
In this letter I will address another possibility, namely the creation of a population of cold, degenerate
dwarfs with masses $\sim 0.1-0.3 \rm M_{\odot}$, which do not burn hydrogen by virtue of their unusual construction.

In section~\ref{construction} I will describe the formation and evolution of said dwarfs  
and the possibilities for their detection.
In section~\ref{genesis} I will present a cosmological context for their formation.

\section{Building a bigger dwarf}
\label{construction}

Traditionally, star formation occurs via the inside-out collapse of a gas cloud in which radiative
cooling and ambipolar diffusion remove the pressure support, allowing rapid contraction to stellar
dimensions. What determines the cessation of accretion (and thus the final mass) is a matter of some debate, be it
the onset of nuclear fusion, competition for gaseous resources or back-reaction from protostellar outflows.
At the end of this contraction, the protostar ignites nuclear burning in the core if the density and
temperature are high enough. Otherwise it becomes a degeneracy-supported brown dwarf and cools rapidly to
invisibility. The dividing line between these outcomes is a mass of $\sim 0.08 \rm M_{\odot}$ (Burrows et al 1993).

However, Lenzuni, Chernoff \& Salpeter (1992) \& Salpeter (1992) demonstrated that one could raise the
hydrogen burning limit by accreting material onto a brown dwarf ($\sim 0.01 \rm M_{\odot}$) at low enough
($\sim 10^{-11}-10^{-9} \rm M_{\odot} yr^{-1}$) rates that the material settles onto the dwarf with
low entropy. In principle, this procedure is limited only by the pyconuclear burning rate, and the 
hydrogen burning limit could be raised to $\sim 1 \rm M_{\odot}$. However, the small but finite entropy of the
accreted material is likely to limit this mass to somewhat less than this
(Salpeter (1992) estimates a mass limit $\sim 0.15 \rm M_{\odot}$).
The limiting entropy of accreted metal-free material in spherical geometry is determined by the opacity minimum in the settling layer
above the star, at $\sim 3000$~K  where the competition between $\rm H^-$ and $\rm H_2$
opacity is approximately equal (Lenzuni et al 1992). The addition of metals will raise the opacity and will eventually
lower
the hydrogen burning limit to original levels. On the other hand, the accretion of material from a disk
(possible if the accretion is from an inhomogeneous medium) may serve to increase the limiting mass,
depending on conditions at the inner edge of the disk.

To explore this scenario further than the semi-analytic analysis of Salpeter, I have constructed a sequence
of brown dwarf models using the same atmospheric and evolution codes used to describe old white dwarfs
(Hansen 1998, 1999) while incorporating the degeneracy corrections to the nuclear burning from Salpeter (1992).
I do not model the accretion process onto the dwarf as done by Lenzuni et al (1992). Given the uncertainty in
the state of material accreted from a disk, I prefer to address the question in another fashion, namely; how
hot can a dwarf of given mass be before nuclear burning turns it into a star?
For each mass, I have constructed a cooling sequence without including nuclear burning. I then consider
the effect of switching on nuclear burning at various points on the cooling sequence. For each mass there
is a transition point on the sequence above which there is a slow runaway and the dwarf becomes a normal
hydrogen burning star. Below the transition, nuclear burning is not strong enough to overwhelm the cooling
luminosity and the star fades as a brown dwarf. Thus, I determine the range of parameter space which can
accommodate brown dwarfs of varying mass. This represents the range of dwarf configurations which can
potentially be constructed by the slow accretion of low entropy material.\footnote{Deuterium burning during
construction may change the adiabat for a given object, but is not strong enough to change this bound (Salpeter 1992)}

The first important point to note is that, for dwarfs of primordial composition, the normal hydrogen burning limit is
raised to $\sim 0.1 \rm M_{\odot}$ as noted by Nelson (1989), a consequence of the change in boundary
condition resulting from the lower atmospheric opacity. As the mass increases, the transition temperature
drops rapidly as we pass through the range of masses 0.13-0.17$\rm M_{\odot}$. The effective temperatures for
models at the transition point are $\sim 1500$~K.
At masses $> 0.17 \rm M_{\odot}$, electron degeneracy in the centre leads to formation of an isothermal core
where energy transport by electron conduction dominates convection. This results in lower central temperatures
than the adiabatic case and the effective temperature at the transition point remains essentially constant
in the range $\sim 1200-1500$~K for masses $0.17 - 0.3 \rm M_{\odot}$.
Thus, if we allow material to be accreted with entropies appropriate to $T_{\rm eff} \sim 1500$~K
atmospheres, the mass limit can be raised considerably above the $0.15 \rm M_{\odot}$ estimated by Salpeter (1992).
These stars represent a configuration containing elements of both brown dwarfs (in terms of composition and lack of
 nuclear burning) and
white dwarfs (in terms of mass and internal structure). Thus, for the purposes of clarity later, I shall refer to 
these ($>0.1 \rm M_{\odot}$) objects as `Beige dwarfs'.

Once formed, these dwarfs will fade slowly, radiating what little thermal energy they possess, just as brown
and white dwarfs do.
The cooling of the various models is shown in Figure~\ref{HR}. The watershed nature of the $0.15 \rm M_{\odot}$ model is apparent.
 For smaller masses, the evolution is similar to that of
a traditional brown dwarf, with a  rapid fading to insignificance within 5-6~Gyr.
 The $0.15 \rm M_{\odot}$ model retains a significant contribution from
nuclear burning for several~Gyr after birth, so that it takes longer to cool and therefore is the brightest
model after $\sim 15 \rm Gyr$. For more massive models, the central transition temperatures are low enough that the dwarfs essentially
begin life in a cool, white dwarf-like configuration and do not cool significantly further within a Hubble time. Also shown is
a 0.6~$\rm M_{\odot}$
Carbon/Oxygen white dwarf with hydrogen atmosphere from Hansen (1999). Thus, the larger radii of the beige dwarfs
mean they are $\sim 1$~magnitude brighter (since they have similar effective temperatures to the white dwarfs). For
masses $\sim 0.3 \rm M_{\odot}$ the beige dwarfs 
 will be superficially similar in appearance to white dwarfs.
 Figure~\ref{HR} shows
the colours calculated for the HST WFPC bandpasses of Holtzmann et al (1995).
 Furthermore, a population of such objects can be distinguished from a white dwarf population by the very
different cooling sequence. Beige dwarfs are born with effective temperatures $\sim 1500-3000$~K, in which molecular
hydrogen is already a strong source of opacity (e.g. Borysow, Jorgensen \& Zheng 1997), so that they cannot populate a region equivalent to the upper part of
the white dwarf cooling sequence (i.e. the initial redward evolution of the white dwarf track), which occurs for temperatures
$>4000$~K.

\section{Cosmological Considerations}
\label{genesis}

This particular mode of star creation requires fairly special conditions to be important. The accretion rate must be in a narrow range
$\sim 10^{-11}-10^{-9} \rm M_{\odot}.yr^{-1}$ to yield both significant mass accretion while allowing the 
accreted material to retain only low entropies (Lenzuni et al 1992).
 Furthermore, such rates are well above the $\sim 10^{-18} \rm M_{\odot} yr^{-1}$
experienced in the local ISM, implying that such rates must be related to the initial conditions for the formation
of such objects.

Let us assume that brown dwarfs form in small clusters. Unless the process is highly efficient, there will be
a substantial amount of gas present as well.
For a cluster of mass M and radius R, the
ambient density $\rho \sim 0.25 M/R^3$ and velocity $V^2 \sim G M/R$ yield a Bondi-Hoyle accretion rate for a 
$0.01 \rm M_{\odot}$ brown dwarf of
\be
\dot{M} \sim 6.4 \times 10^{-11} {\rm M_{\odot} yr^{-1}} \left( \frac{M}{10^2 \rm M_{\odot}} \right)^{-1/2} 
\left( \frac{R}{0.1 \rm pc} \right)^{-3/2}
\ee
Such an estimate lies within the acceptable range for transformation of brown dwarfs into beige dwarfs.
 But we need to know what size clusters and collapsed objects do we expect from primordial star formation.

The formation of primordial stars is intricately tied to the cooling mechanisms of primordial gas and thus to the formation of molecular hydrogen,
the dominant coolant in dense, metal-free media. In the traditional cosmological scenario of hierarchical gravitational collapse,
gas falls into  the potential well of a cold, non-baryonic dark matter component, is shock-heated to virial temperatures and cools to form the baryonic component
of protogalaxies. 
Fall \& Rees (1985) identified a thermal instability in such hot gaseous halos, in which the gas fragments into cold clumps surrounded
by  a hot, high pressure ambient medium. The isobaric collapse of the cold clumps was assumed to halt at $\sim 10^4$~K where the 
Lyman edge leads to inefficient cooling. The characteristic mass of such isobaric collapsing clumps was determined to be $\sim 10^5-10^6 \rm M_{\odot}$ and was thereby identified as a possible mechanism for forming globular clusters. However, non-equilibrium calculations of the
cooling of shock-heated gas (Shapiro \& Kang 1987 and references therein) indicate that cooling progresses faster than recombination,
which leads to a residual electron fraction well above equilibrium levels, thereby promoting the formation of $\rm H_2$ and enhancing the
cooling rate. 
 The result of this process is that the isobaric collapse is able to continue down to temperatures $\sim 30-10^2$~K. This
conclusion is robust unless there is a strong, pre-existing source of ultra-violet radiation (such as an AGN) in
 the same protogalaxy (Kang et al 1990), i.e. cooling below $10^4$~K cannot be stopped by a UV background alone.
The Bonner-Ebert critical mass (Ebert 1955; Bonner 1956) for gravitational instability in high pressure media is
\be
M_{crit}  =  1.18 \left( \frac{k_b T_{\rm cool}}{\mu m_{\rm p}} \right)^2 G^{-3/2} p^{-1/2} 
\ee
where $p \sim n_{\rm hot} k T_{\rm vir}$ is the pressure in the hot, virialised gas phase. The density of the hot phase is such
that the cooling time is comparable to the dynamical time (Fall \& Rees 1985), so that the pressure is determined entirely by the 
virial temperature and cooling function.
For virial temperatures appropriate to our galactic halo ($\sim 10^6$~K) the critical mass clusters will range from
$\sim 10^2-10^3 \rm M_{\odot}$, depending on the final cooled temperature (assuming a variation between 30-100~K).
The appropriate accretion rate is thus
\be
\dot{M} \sim  1.6 \times 10^{-11} {\rm M_{\odot} yr^{-1}} \left( \frac{M_{\rm bd}}{10^{-2} \rm M_{\odot}} \right)^2  
\left( \frac{T_{\rm cool}}{\rm 30~K} \right)^{-5/2} 
\ee
where we have assumed a velocity dispersion for the clump appropriate to the cooled gas temperature and
$M_{\rm bd}$ is the mass of the accreting object. Note that
the Bonner-Ebert mass in this situation is an lower limit on the collapsed mass. Larger clumps with lower
accretion rates can form also, depending on the spectrum of perturbations in the hot gas phase. Nevertheless,
this is exactly the kind of accretion rate we require, low enough to allow accretion of low entropy material but
high enough to allow significant mass accretion on cosmological timescales.

Figure~\ref{mdot} shows the expected accretion rates in such cold clumps as a function of halo velocity dispersion
(i.e. the virial temperature of the hot confining medium). There is some variation allowed depending on the final
temperature to which the clumps can cool and thus the fraction of molecular hydrogen is important. The diagram may be split
into three parts, depending on the accretion rate. For rates $> 10^{-9} \rm M_{\odot} yr^{-1}$, the accreted material
is too hot for the dwarf to remain degenerate and normal hydrogen burning stars in the mass range $\sim 0.1-0.2 \rm M_{\odot}$
are formed (Lenzuni et al 1992). For rates $< 10^{-11} \rm M_{\odot} yr^{-1}$, brown dwarfs will not accrete enough mass
to change their character much in $10^9$~yrs, and so any brown dwarfs formed in such clusters will remain true brown
dwarfs. However, between those two extremes, the conditions are appropriate for the transformation of brown dwarfs
into the beige dwarfs described above in section~\ref{construction}. Furthermore, these conditions are applicable in
the range of $100-300 \rm km.s^{-1}$ that describe the dark matter haloes of galaxies.
Such conditions may also apply in the case of pregalactic cooling flows (Ashman \& Carr 1988; Thomas \& Fabian 1990).
If one wished to extend this picture to the case of cluster cooling flows (e.g. Fabian 1994), the larger virial temperatures and higher pressures
suggest higher accretion rates and thus low mass star formation rather than beige dwarfs.
Thus, it appears that the formation of small gas clusters of appropriate density is a generic feature of gas collapse in CDM haloes
at moderate to high redshifts. However, there is also a requirement that brown dwarfs be the most abundant initial collapsed object. Once again,
the physics of $\rm H_2$ cooling in continued collapse provides the characteristic mass scale, a Jeans mass $< 0.1 \rm M_{\odot}$ 
(Palla, Salpeter \& Stahler 1983). Although the complexity of star formation prevents a conclusive answer, the preceding provide
 plausible conditions for our scenario, namely the copious production of brown dwarf mass objects in relatively dense
media in which they may grow on timescales $\sim 10^9 \rm yrs$.

How long will such accretion episodes last? This is an important question, because Bondi-Hoyle accretion $\propto M_{\rm bd}^2$ and
is thus a runaway process. An obvious concern is that, if gas is too abundant, even accretion at rates initially $<10^{-9} \rm M_{\odot} yr^{-1}$
will eventually lead to sufficient accretion to create a star. The abundance of ambient gas will depend on the efficiency
with which one forms brown dwarfs initially. If the efficiency of conversion of gas into $\sim 0.01 \rm M_{\odot}$ objects is $\sim 10 \%$
then few objects will be able to grow by more than a factor of 10, providing a natural limiting mechanism. Furthermore,
the formation of copious collapsed objects in a small cluster will result in two body relaxation on timescales 
\be
t_{\rm rel} \sim  10^8 {\rm yr} \, \epsilon^{-1} \left( \frac{M_{\rm bd}}{10^{-2} \rm M_{\odot}} \right)^{-1} M_{2}^2 \left( \frac{T_{\rm cool}}{\rm 30~K} \right)^{-3/2}
\label{relax}
\ee
where $\epsilon$ (in units of 0.1 here) is the efficiency of conversion of gas into $0.01 \rm M_{\odot}$ bound objects and
$M_2$ is cluster mass in units of 100~$\rm M_{\odot}$. In fact,
$\epsilon$ and $M_{\rm bd}$ may be regarded as dynamically evolving quantities as the characteristic collapsed object mass grows
through accretion. Although cluster evaporation takes place on timescales $\sim 300 t_{\rm rel}$ (e.g. Spitzer 1987), as more gas
mass is incorporated into dwarfs, the cluster potential becomes `lumpier' and two-body relaxation accelerates, so that
the cluster disruption time is $\sim 5 t_{\rm rel}$ as defined in (\ref{relax}) by the time all the gas mass has been accreted
onto the original seeds.
 This is also approximately equal to
the timescale for Bondi-Hoyle accretion runaway to infinite mass (this is not surprising, given that both accretion and two-body
relaxation are intimately related to the gravitational focussing cross-section). Thus there is a finely
balanced competition between mass accretion and cluster evaporation that may produce a very different mass spectrum than
is usually assumed for primordial objects.

\section{Conclusion}

In this paper I have considered the possibility that thermal instabilities in primordial gas collapse favour the creation of small ($\sim 10^2-10^3 \rm M_{\odot}$)
gaseous clusters, which are ideal sites for the formation of massive brown (a.k.a. beige) dwarfs 
 by the process of Lenzuni, Chernoff \& Salpeter (1992).
This offers a scenario for the formation of baryonic dark matter which meets the requirements of both the microlensing survey mass
limits 
 and those of limited chemical pollution of the interstellar medium and
production of extragalactic light (two stringent constraints on the currently fashionable white dwarf scenario). The scenario
predicts that baryonic dark matter lies predominantly in beige dwarfs $\sim 0.1-0.3 \rm M_{\odot}$
 with probably some
contribution from low mass stars ($\sim 0.2-0.3 \rm M_{\odot}$) as well. This mechanism is also peculiar to the early universe 
because it will become less efficient at constructing beige dwarfs once the accreted
material contains significant metallicity (since greater opacity means material is accreted with more entropy). 
Thus, any  present day analogue is more likely to produce low mass stars.
As such, this 
scenario may also naturally account for the red stellar haloes
of galaxies such as NGC~5907 (Sackett et al 1994). Indeed, Rudy et al (1997) find that the peculiar colours of this halo
requires a population rich in red dwarfs ($<0.25 \rm M_{\odot}$), but of approximately solar metallicity (primordial metallicity
stars are not red enough to explain these colours).

Given the many uncertainties inherent in discussing primordial star formation and galaxy evolution from first principles,
I have also constructed preliminary models  for the evolution and appearance of such objects. Hopefully, deep proper
motion surveys will be able to
 constrain this scenario directly. In particular, such beige dwarfs will be somewhat brighter than white dwarfs and should
also occupy a restricted region of the Hertzsprung-Russell diagram, i.e. they will not display a cooling track like a white
dwarf population would.

\clearpage


\figcaption[hr.ps]{The solid line represents a 0.6~$\rm M_{\odot}$ white dwarf cooling track.
The dotted lines are the cooling tracks of degenerate brown dwarfs of mass 0.13, 0.15, 0.17,
 0.2, 0.25 and 0.3 $\rm M_{\odot}$. Each curve has three open circles plotted on it, at ages of 5~Gyr,
10~Gyr and 15~Gyr. The 0.13~$\rm M_{\odot}$ object cools to invisibility in less than 10~Gyr. Note that these
represent the cooling of the hottest possible models only, i.e. cooler models are possible for each mass.
 \label{HR}}

\figcaption[mdot.ps]{  Here we find the accretion rate we expect for brown dwarfs formed in clusters 
 of cold gas resulting from the thermal instability of shock heated gas in dark matter haloes (Fall \& Rees 1985) of
varying velocity dispersion and for various final temperatures of the cold phase (which will depend on the amount
of $\rm H_2$ that is formed). The dotted lines delineate the various outcomes, depending on the
accretion rate. The two lower lines require significant accretion on timescales of $1$~Gyr and $10$~Gyr
respectively. The dashed line indicates a velocity dispersion of $50$~km/s. For smaller haloes, there
is little non-equilibrium production of $\rm H_2$ behind virialisation shocks (Shapiro \& Kang 1987),
but the accretion rates are very low anyway for such small haloes.
  \label{mdot}}

\clearpage


\begin{references}
\reference{MACHO1} Alcock, C. et al, 1993, Nature, 365, 621
\reference{AMACHO} Alcock, C. et al, 1997, ApJ, 486, 697
\reference{AC} Ashman \& Carr, 1988, MNRAS, 234, 219
\reference{Au1} Aubourg, E. et al, 1993, Nature, 365, 623
\reference{Au2} Aubourg, E. et al, 1995, A\&A, 301, 1
\reference{BFGK}  Bahcall, J. N., Flynn, C., Gould, A. \& Kirhakos, S., 1994, ApJ, 435, L51
\reference{BE} Bonner, W. B., 1956, MNRAS, 116, 351
\reference{BJZ} Borysow, A., Jorgensen, U. G. \& Zheng, C., 1997, A\&A, 324, 185
\reference{BHSL} Burrows, A., Hubbard, W. B., Saumon, D. \& Lunine, J. I., 1993, ApJ, 406, 158
\reference{CL} Carr, B. J., \& Lacey, C. G., 1987, ApJ, 316, 23
\reference{DIJR} De Paolis, F., Ingrosso, G., Jetzer, P. \& Roncadelli, M., 1998, ApJ, 500, 59
\reference{Eb} Ebert, R., Zs. Astr., 37, 217
\reference{Fab} Fabian, A. C., 1994, ARA\&A, 32, 277
\reference{FR} Fall, S. M. \& Rees, M. J., 1985, ApJ, 298, 18
\reference{GM} Gibson, B. \& Mould, J., 1997, ApJ, 482, 98
\reference{GF} Graff, D. S. \& Freese, K., 1996, ApJ, 456, L49
\reference{Han} Hansen, B. M. S., 1998, Nature, 394, 860
\reference{Han99} Hansen, B. M. S., 1999, ApJ, in press, astro-ph/9903025
\reference{Holtz} Holtzman, J. A., et al, 1995, PASP, 107, 1065
\reference{KSFR} Kang, H., Shapiro, P. R., Fall, S. M. \& Rees, M. J., 1990, ApJ, 363, 488
\reference{K97} Kerins, E., 1997, A\&A, 328, 5
\reference{KE} Kerins, E. \& Evans, N. W., 1998, ApJ, 503, L75
\reference{LCS2} Lenzuni, P., Chernoff, D. F. \& Salpeter, E. E., 1992, ApJ, 393, 232
\reference{Nel} Nelson, L. R., 1989, in {\em Baryonic Dark Matter}, ed. D. Lynden-Bell \& G.Gilmore,
NATO ASI Series, Kluwer,  pg 67
\reference{PSS} Palla, F., Salpeter, E. E. \& Stahler, S. W., 1983, ApJ, 271, 632
\reference{RWHFH} Rudy, R. J., Woodward, C. E., Hodge, T., Fairfield, S. W. \& Harker, D. E., 1997, Nature, 387, 159
\reference{SMHB} Sackett, P. D., Morrison, H. L., Harding, P. \& Boroson, T. A., 1994, Nature, 370, 441
\reference{Salt} Salpeter, E. E., 1992, ApJ, 393, 258
\reference{SK} Shapiro, P. R. \& Kang, H., 1987, ApJ, 318, 32
\reference{Spit} Spitzer, L., 1987, Dynamical Evolution of Globular Clusters, Princeton University Press, Princeton, New Jersey
\reference{TF} Thomas \& Fabian, 1990, MNRAS, 246, 156
\end{references}
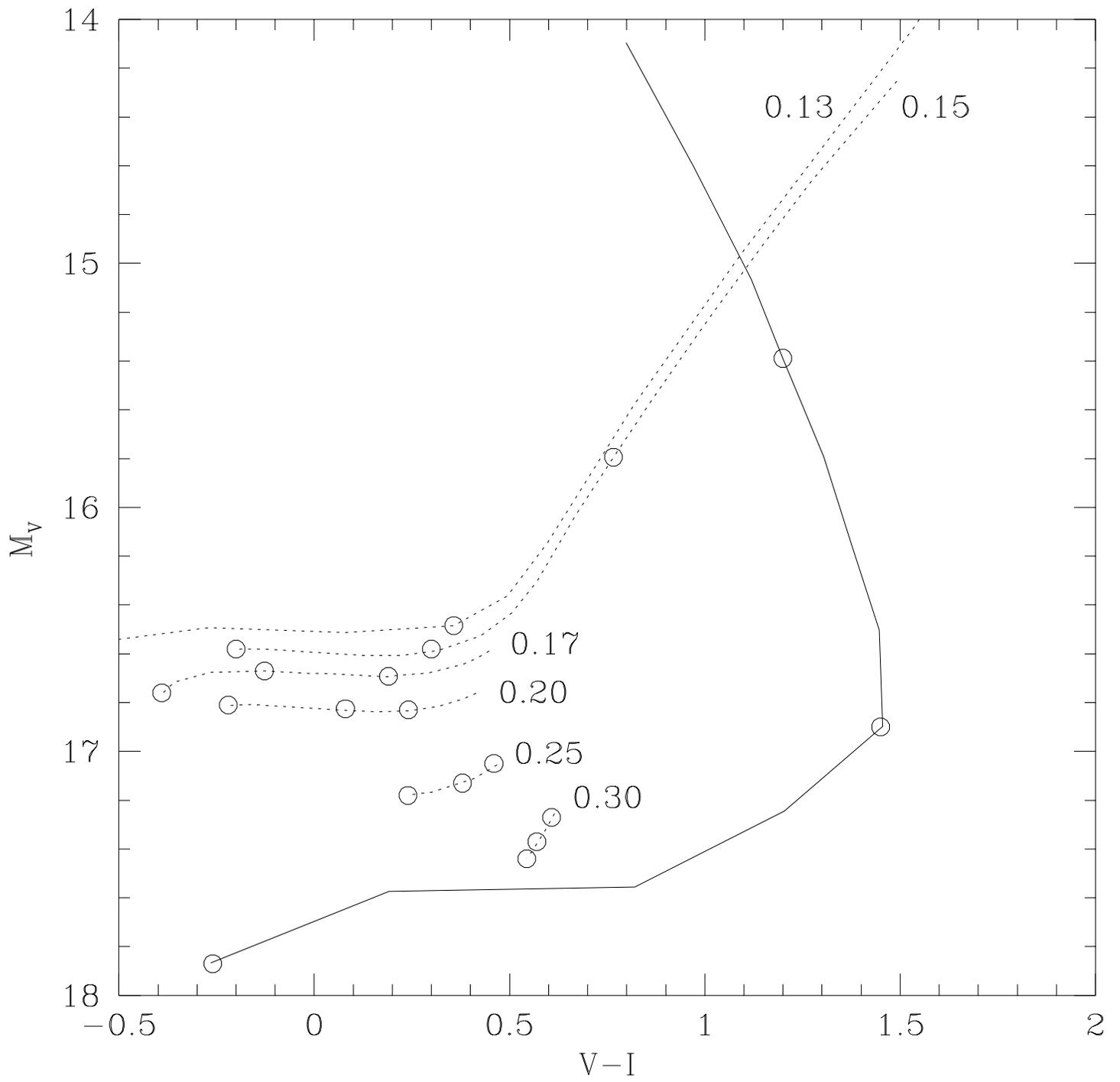
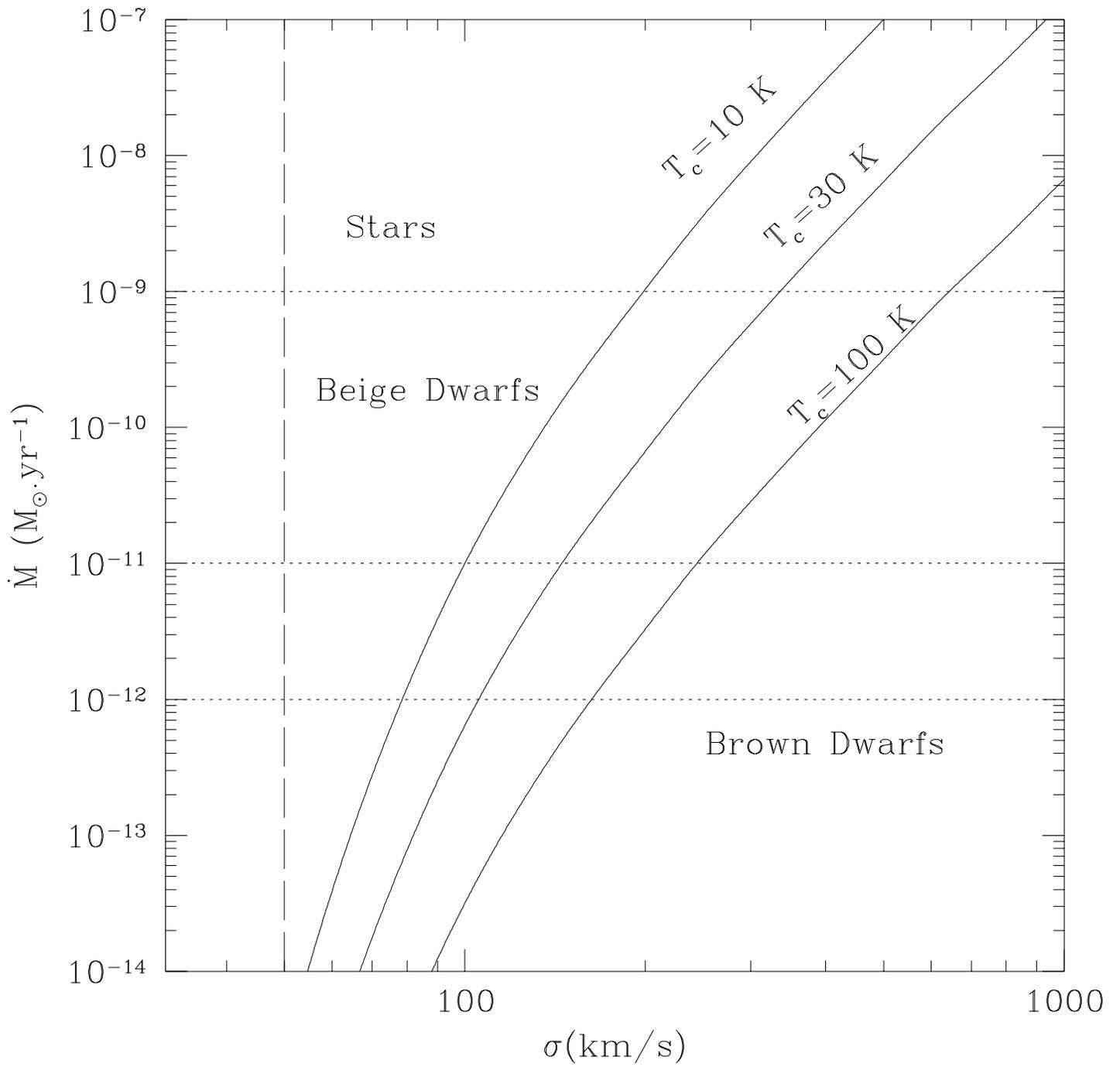
\end{document}